\documentstyle[aps,preprint,epsf]{revtex} 
\begin{document} 
\widetext
 
\title{Continuous canonical transformation \\
       for the double exchange model}

\author{ T.~Doma\'nski}
\address{Institute of Physics, 
         M.~Curie Sk\l odowska University,
         20-031 Lublin, Poland} 
\date{\today}  
\maketitle 
\draft 

\begin{abstract}
The method of continuous canonical transformation
is applied to the double exchange model with a purpose
to eliminate the interaction term responsible for
non conservation of magnon number. Set of differential
equations for the effective Hamiltonian parameters
is derived. Within the lowest order (approximate) 
solution we reproduce results of the standard 
(single step) canonical transformation. Results 
of the selfconsistent numerical treatment are 
compared with the other known studies for this 
model. 
\end{abstract}

\vspace{1cm}

\section{Introduction}

The discovery of colossal magnetoresistance in 
the doped manganites attracted new interests 
in studying itinerant ferromagnetism phenomena. 
The double exchange model (DEX), introduced a long 
time ago \cite{Zener-51,DEXstudies}, seems to be 
a good starting point to explain the paramagnetic 
- ferromagnetic transition. Conduction $e_{g}$ 
electrons interact via the strong Hund's coupling
with the localized Mn ions (their spin being 
$S=3/2$). This interaction drives the core
Mn spins to ferromagnetic alignment, owing to 
the kinetic processes of itinerant electrons.
Whether the DEX scenario itself is enough for 
an explanation of magnetoresistance in manganites, 
or should be supplemented by other realistic 
effects like e.g.\ lattice Jahn-Teller distortions 
\cite{Millis-95}, competition with the superexchange
\cite{Shannon-00,Perkins-99} and strong Coulomb 
interactions between electrons \cite{Feinberg-00}, 
is an open question under discussion.

In this paper we want to reexamine physics of
the DEX model using the renormalization group 
approach in a version proposed recently by Wegner 
\cite{Wegner-94} and independently by G\l azek 
and Wilson \cite{Glazek-94}. This procedure known 
as the {\em flow equation method} has
proved to be a powerful tool, especially for 
analysis of models composed of several coupled
subsystems. So far, it has been successfully
applied to the problem of electron - phonon 
interaction \cite{el-ph} (where the Fr\"ohlich 
transformation has been revised), the single
impurity Anderson model \cite{SIAM}, the spin 
- boson model for dissipative systems \cite{spin-boson}, 
the spin - polaron coupling for $t-J$ model
\cite{spin-polaron}, DEX model in the RKKY 
(small coupling) limit \cite{RKKY}, and the charge 
exchange interaction for the boson - fermion model 
\cite{Domanski-00}. The same method has also been
used for studying effects of correlations, e.g.
in the Hubbard model \cite{tU} ($t/U$ expansion),
the large spin Heisenberg Hamiltonian \cite{largeS}
($1/S$ expansion), etc. Recently there has been
proposed a highly sophisticated {\em computer aided
perturbation method} based on the flow equation 
technique to study the dimerized spin models 
\cite{Uhrig}. It is our belief, that the flow equation 
technique can be a reliable source of information 
also for the DEX model. In particular we would
like to consider the strong (ferromagnetic) coupling 
limit (relevant for manganites) and compare our 
results to other studies of this model. 

The DEX model for ferromagnetism is described by 
the Kondo type Hamiltonian \cite{Kasuya-56}
\begin{equation}
H = \sum_{{\bf k},\sigma} \left( \varepsilon_{\bf k} -
\mu \right) c^{\dagger}_{{\bf k}\sigma} c_{{\bf k}\sigma}
- J_{H} \sum_{i,\sigma,\sigma'} \left( {\bf S}_{i}  
{\bf s}_{i} \right)^{\sigma \sigma'}
c_{i\sigma}^{\dagger} c_{i\sigma'}
\label{Kondo}
\end{equation} 
where $c_{{\bf k}\sigma}^{(\dagger)}$ correspond to annihilation
(creation) operators of the conduction $e_{g}$ electrons,
${\bf s}_{i}$ denotes their spin operators and ${\bf S}_{i}$
stands for the spin operator of Mn ions. Local ferromagnetic
interaction is characterized by the Hund's coupling $J_{H}>0$,
for manganites known to be very large. With a help of the 
Holstein-Primakoff transformation we can represent the spin 
operators ${\bf S}_{i}$ via the magnon operators 
$a_{i}^{\dagger}$, $a_{i}$ such that $S^{-}_{i} = 
a_{i}^{\dagger}(2S-a^{\dagger}_{i} a_{i})^{1/2}$, 
$S^{+}_{i}=(S^{-}_{i})^{\dagger}$, $S^{z}_{i}=
S-a^{\dagger}_{i}a_{i}$. Magnon operators $a^{\dagger}_{i}$, 
$a_{i}$ obey the boson (commutation) relations. 

At sufficiently low temperatures the system is close to
ferromagnetic (ground state) ordering, so we can simplify
spin operators to $S^{-}_{i} \simeq a_{i}^{\dagger} \sqrt{2S}$,
$S^{+}_{i} \simeq a_{i} \sqrt{2S}$. Using the Pauli
operators for $e_{g}$ electron spins one gets the model
Hamiltonian in a following form
\begin{eqnarray}
H & = & \sum_{{\bf k},\sigma} \xi^{\sigma}_{\bf k}
c^{\dagger}_{{\bf k} \sigma}c_{{\bf k}\sigma}
+ \frac{J_{H}}{2N} \sum_{{\bf q},{\bf p},{\bf k}}
a^{\dagger}_{{\bf p}+{\bf q}}a_{\bf p} \left(
c^{\dagger}_{{\bf k}-{\bf q}\uparrow}c_{{\bf k}\uparrow}
- c^{\dagger}_{{\bf k}-{\bf q}\downarrow}
c_{{\bf k}\downarrow} \right) \nonumber \\
& - & J_{H}
\sqrt{\frac{S}{2N}} \sum_{{\bf q},{\bf k}}  \left(
a^{\dagger}_{\bf q} c^{\dagger}_{{\bf k}-{\bf q}\uparrow}
c_{{\bf k} \downarrow} + {\rm h.c.} \right) \;,
\label{model}
\end{eqnarray}
with $\xi^{\sigma}_{\bf k}=\varepsilon_{\bf k}-\mu^{\sigma}$
and $\mu^{\uparrow}=\mu+\frac{1}{2}SJ_{H}$, $\mu^{\downarrow}
=\mu-\frac{1}{2}SJ_{H}$. If temperature is comparable with
$J_{H}/k_{B}$ then one should include the higher order terms
in the square root expansion $\sqrt{1-(a_{i}^{\dagger}a_{i}
/2S)}$ for magnon operators.

In what follows, we design the unitary transformation 
to eliminate last part of the Hamiltonian (\ref{model})
linear in $a_{q}^{(\dagger)}$ operators, which is
responsible for a violation of the magnon number 
$\sum_{i} <a_{i}^{\dagger}a_{i}>$. To the leading 
order of $1/S$ one can eliminate this exchange interaction
using a single step canonical transformation $e^{A}He^{-A}$ 
with the generating operator $A$ given by 
\cite{Nagaev-98}
\begin{equation}
A = J_{H}\sqrt{\frac{S}{2N}} \sum_{{\bf k},{\bf q}} 
\left( \frac{a_{\bf q}^{\dagger}c_{{\bf k}-{\bf q}\uparrow}
^{\dagger} c_{{\bf k}\downarrow}}{\xi_{{\bf k}-
{\bf q}}^{\uparrow}-\xi_{\bf k}^{\downarrow}} - 
\mbox{h.c.} \right) \;. \label{standard}
\end{equation}
However, this transformation generates a whole lot 
of higher order interactions, their amplitudes being  
eventually not negligible. Recently it has been shown 
by means of the standard perturbation treatment 
\cite{Golosov-00} that these interactions give rise 
to quantum corrections for spin wave spectrum 
(both for dispersion and the life time effects).  
That such corrections are in fact important it is
known independently from the direct numerical studies
of finite chains \cite{Kaplan-97} and from analysis 
based on the variational wave functions \cite{Wurth-98}.

Instead of the single step transformation 
(\ref{standard}) we propose in this paper a different
method using an infinite sequence of the infinitesimal 
transformations what gives us more control for a 
derivation of the required effective Hamiltonian.
In particular, we want the higher order many-body 
interactions to be as small as possible, thus being 
more tractable via the standard perturbation study.

In the following section we give a brief introduction 
to the method of continuous transformation and derive 
the corresponding flow equations for parameters of
the DEX model. Next, we discuss an analytical 
approximate solution of these equations and 
compare it with results of the standard single 
step transformation. In the last part we present
the selfconsistent numerical solution for the
model parameters along with some rough estimation 
of the spin stiffness coefficient. 

\section{Formulation of flow equations}

A main idea of the method is to transform the initial
Hamiltonian $H$ through the series of unitary transformations
$H(l)=U(l)HU^{\dagger}(l)$, labeled with some continuous  
{\em flow parameter} $l$. In a course of transformation
the Hamiltonian evolves according to the following 
{\em flow equation}
\begin{eqnarray}
\frac{dH(l)}{dl} = [ \eta(l), H(l) ] \;,
\label{equation}
\end{eqnarray}
where the generator $\eta(l)$ is related to $U(l)$ via 
$\eta(l)=(dU(l)/dl)U^{\dagger}(l)$. This operator has to
be chosen depending on a purpose of the transformations. 
Wegner has shown \cite{Wegner-94} that with $\eta(l)=
[H(l)-H_{int}(l),H(l)]$ one can eventually eliminate 
the (perturbation) part of the Hamiltonian $H_{int}(l 
\rightarrow \infty ) = 0$, provided that no degenerate 
states are encountered. Alternative choice for $\eta$, 
efficient even in a presence of degeneracies, has been 
proposed recently by Mielke \cite{Mielke-98}. In this 
work we use the slightly modified Wegner's proposal
for $\eta(l)$.

To be specific, we define the interaction part as
\begin{eqnarray}
H_{int}(l) = \frac{-1}{\sqrt{N}} \sum_{{\bf k},{\bf q}} 
\left( I_{{\bf k},{\bf q}}(l) a_{\bf q}^{\dagger} 
c_{{\bf k}-{\bf q},\uparrow}^{\dagger} 
c_{{\bf k}\downarrow} + \mbox{h.c.} \right) \;.
\label{perturb}
\end{eqnarray}
The remaining part $H_{0}(l) = H(l)-H_{int}(l)$ may 
contain not only the other two terms of (\ref{model}) 
but additionally also contributions induced by 
transformation for $l>0$. To take these into account 
we assume $H_{0}(l)$ to have the following structure
\begin{eqnarray}
H_{0}(l) & = & \sum_{{\bf k},\sigma}
\xi_{\bf k}^{\sigma}(l) c_{{\bf k}\sigma}^{\dagger}
c_{{\bf k}\sigma} + \frac{1}{N} 
\sum_{{\bf k},{\bf k}',{\bf q},{\bf q}'} 
\delta_{{\bf k}+{\bf q},{\bf k}'+{\bf q}'} \left[
U_{{\bf k},{\bf q},{\bf q}',{\bf k}'}(l) c_{{\bf k}
\downarrow}^{\dagger}c_{{\bf q}\uparrow}^{\dagger}
c_{{\bf q}'\uparrow} c_{{\bf k}'\downarrow}
\right. \nonumber \\ & + & \left. \left(
M_{{\bf k},{\bf k}',{\bf q},{\bf q}'}^{\uparrow}(l) 
c^{\dagger}_{{\bf k}\uparrow} c_{{\bf k}'\uparrow} -
M_{{\bf k},{\bf k}',{\bf q},{\bf q}'}^{\downarrow}(l)
c^{\dagger}_{{\bf k}\downarrow} c_{{\bf k}'\downarrow} 
\right) a_{\bf q}^{\dagger}a_{{\bf q}'}  
\right] 
+ \delta H_{0}(l) \;,
\label{diag}
\end{eqnarray}
where $\delta H_{0}(l)$ contains all types of 
interactions not shown explicitly in (\ref{diag}). 
The initial conditions for the model parameters 
read
\begin{eqnarray}
\xi_{\bf k}^{\sigma}(0) = \varepsilon_{\bf k} - 
\mu^{\sigma} \;, & \hspace{1.5cm} & 
U_{{\bf k},{\bf q},{\bf q}',{\bf k}'}(0) = 0 \;, 
\label{condition1} \\
I_{{\bf k},{\bf q}}(0) = J_{H} \sqrt{\frac{S}{2}} \;,  & & 
M_{{\bf k},{\bf k}',{\bf q},{\bf q}'}^{\sigma}(0) = 
\frac{J_{H}}{2} \;.
\label{condition2}
\end{eqnarray}
We choose the generating operator 
$\eta(l)=[\sum_{{\bf k},\sigma} \xi_{\bf k}^{\sigma}(l) 
c_{{\bf k}\sigma}^{\dagger} c_{{\bf k}\sigma},H_{int}(l)]$
which explicitly is given by
\begin{eqnarray}
\eta(l) = \frac{-1}{\sqrt{N}} \sum_{{\bf k},{\bf q}} 
\alpha_{{\bf k},{\bf q}}(l) \left( I_{{\bf k},{\bf q}}(l) 
a_{\bf q}^{\dagger}c_{{\bf k}-{\bf q},\uparrow}^{\dagger}
c_{{\bf k}\downarrow} - \mbox{h.c.} \right) ,
\label{generator}
\end{eqnarray}
where $\alpha_{{\bf k},{\bf q}}(l)=\xi_{{\bf k}-{\bf q}}
^{\uparrow}(l)-\xi_{\bf k}^{\downarrow}(l)$. Notice, that 
(\ref{generator}) has similar structure to the generating
operator $A$ of the standard transformation (\ref{standard}).

Using the general flow equation (\ref{equation}) we obtain
\begin{eqnarray}
\frac{dH(l)}{dl} & = & \frac{1}{\sqrt{N}} 
\sum_{{\bf k},{\bf q}} \left( \alpha_{{\bf k},
{\bf q}}(l) \right)^{2}  \left( I_{{\bf k},
{\bf q}}(l) a_{\bf q}^{\dagger} c_{{\bf k}-
{\bf q},\uparrow}^{\dagger}c_{{\bf k}\downarrow} 
+ \mbox{h.c.} \right) - \frac{2}{N} \sum_{{\bf k},
{\bf q}} \alpha_{{\bf k},{\bf q}}(l) |I_{{\bf k},
{\bf q}}(l)|^{2} c_{{\bf k}\downarrow}^{\dagger} 
c_{{\bf k}\downarrow} 
\nonumber \\ & + &
\frac{1}{N} \sum_{{\bf k},{\bf k}',{\bf q},{\bf q}'} 
\left( \alpha_{{\bf k},{\bf q}}(l) + \alpha_{{\bf k}',
{\bf q}'}(l) \right) I_{{\bf k},{\bf q}}(l) I_{{\bf k}',
{\bf q}'}^{*}(l) \left[ a_{\bf q}^{\dagger}a_{{\bf q}'}
\left( c_{{\bf k}-{\bf q}\uparrow}^{\dagger} 
c_{{\bf k}'-{\bf q}'\uparrow} \delta_{{\bf k},{\bf k}'} 
\right. \right. \nonumber \\ & - & \left. \left.
c_{{\bf k}'\downarrow}^{\dagger}c_{{\bf k}\downarrow}
\delta_{{\bf k}-{\bf q},{\bf k}'-{\bf q}'} \right) 
+ c_{{\bf k}'\downarrow}^{\dagger} c_{{\bf k}-{\bf q}
\uparrow}^{\dagger} c_{{\bf k}'-{\bf q}'\uparrow} 
c_{{\bf k}\downarrow} \delta_{{\bf q},{\bf q}'} \right]
\nonumber \\ & + & [\eta(l),\delta H_{0}(l)] 
+ O( \hspace{0.2mm} I \hspace{0.2mm} M \hspace{0.2mm}) 
+ O( \hspace{0.2mm} I \hspace{0.3mm} U \hspace{0.2mm}) 
\;. \label{flow_hamil}
\end{eqnarray}
Terms of the order $I(l)M(l)$, $I(l)U(l)$ are
symbolically denoted as $O(\hspace{0.2mm}I\hspace{0.2mm}
M\hspace{0.2mm})$ and $O(\hspace{0.2mm}I\hspace{0.3mm}
U\hspace{0.2mm})$. If they were included in the diagonal
part (\ref{diag}) through $\delta H_{0}(l)$ they would 
induce some higher order interactions given by
$[\eta(l),\delta H_{0}(l)]$. 

Equation (\ref{flow_hamil}) is a differential flow
equation for the Hamiltonian which has to be solved.
In the next section we solve it approximately neglecting
the last three terms on the right hand side.

\section{Lowest order solution}

It is instructive to study first the flow equation 
(\ref{flow_hamil}) with the last three terms on the 
right hand side omitted. It means that we neglect the 
interactions expressed by more than four operators,
i.e. scattering between more than two particles.
On a level of this assumption one may expect that 
effective Hamiltonian should differ from its initial 
form (\ref{model}) by some correction comparable with
$H'_{2}$ of the Ref.\ \cite{Golosov-00}. The set of 
the flow equations for parameters of the DEX 
model Hamiltonian is simply given by
\begin{eqnarray}
\frac{dI_{{\bf k},{\bf q}}(l)}{dl} 
& = & - \alpha_{{\bf k},{\bf q}}^{2}(l) 
I_{{\bf k},{\bf q}}(l) \;,  
\label{flow_I} \\
\frac{\xi_{\bf k}^{\downarrow}(l)}{dl} & = & - \; 
\frac{2}{N} \sum_{\bf q} \alpha_{{\bf k},{\bf q}}(l) 
|I_{{\bf k},{\bf q}}(l)|^{2} \;,
\label{flow_xi} \\
\frac{dU_{{\bf k},{\bf p},{\bf p}',{\bf k}'}(l)}{dl} 
& = & \left( \alpha_{{\bf k}',{\bf k}'-{\bf p}}(l) + 
\alpha_{{\bf k},{\bf k}-{\bf p}'}(l) \right)
I_{{\bf k}',{\bf k}'-{\bf p}}(l) I^{*}_{{\bf k},
{\bf k}-{\bf p}'}(l) \;,
\label{flow_U} \\
\frac{M^{\uparrow}_{{\bf k},{\bf k}',{\bf q},{\bf q}'}
(l)}{dl} & = & \left( \alpha_{{\bf k}+{\bf q},{\bf q}}(l) 
+ \alpha_{{\bf k}'+{\bf q}',{\bf q}'}(l) \right)
I_{{\bf k}+{\bf q},{\bf q}}(l) 
I^{*}_{{\bf k}'+{\bf q}',{\bf q}'}(l) \;,
\label{flow_Mup} \\
\frac{M^{\downarrow}_{{\bf k},{\bf k}',{\bf q},
{\bf q}'}(l)}{dl} & = & \left( \alpha_{{\bf k}',{\bf q}}
(l)+\alpha_{{\bf k},{\bf q}'}(l) \right) 
I_{{\bf k}',{\bf q}}(l) I^{*}_{{\bf k},{\bf q}'}(l) .
\label{flow_Mdown} 
\end{eqnarray}
There is no renormalization for spin $\sigma=\uparrow$ 
electrons, so $\xi_{\bf k}^{\uparrow}(l) = \xi_{\bf k}
^{\uparrow}$. 

Still, the set of flow equations (\ref{flow_I}
-\ref{flow_Mdown}) is impossible to solve by 
other than numerical way. To get some insight 
in a process of the continuous transformation 
for $H(l)$ we assume further that the energy 
$\xi_{\bf k}^{\downarrow}(l)$ is only weakly 
affected by renormalization. We verified 
validity of such assumption by solving 
selfconsistently the flow equations numerically 
for one dimensional tight binding electron dispersion 
(see the next section). We thus can drop $l$ 
dependence of the parameter $\alpha_{{\bf k},
{\bf q}}(l)$ on the right hand side of flow 
equations. One easily obtains an analytical 
solution for a flow of the exchange coupling 
\begin{eqnarray}
I_{{\bf k},{\bf q}}(l) = I_{{\bf k},{\bf q}}(0)
e^{ - \alpha_{{\bf k},{\bf q}}^{2}l} =
J_{H}\sqrt{\frac{S}{2}}e^{-(\xi_{{\bf k}-{\bf q}}^{\uparrow}
- \xi_{\bf k}^{\downarrow})^{2}l} \;.
\label{I_approx}
\end{eqnarray}
In the limit $l \rightarrow \infty$ the exchange
coupling drops asymptotically to zero and so does
the part (\ref{perturb}) of the Hamiltonian, 
$H_{int}(l\rightarrow\infty)=0$.

Determination of the other $l$-dependent parameters, 
such as $\xi^{\downarrow}(l)$, $U(l)$ and $M^{\sigma}(l)$ 
is straightforward. We summarize our results by showing 
values of these parameters in the limit $l \rightarrow 
\infty$ 
\begin{eqnarray}
\xi^{\downarrow}_{\bf k}(\infty) & = & \xi^{\downarrow}
_{\bf k} - \frac{J_{H}^{2}S}{2} \; \sum_{\bf q} \frac{1}
{\xi^{\uparrow}_{\bf q} - \xi^{\downarrow}_{\bf k}} \;,  
\label{appr_xi}
\\
U_{{\bf k},{\bf p},{\bf p}',{\bf k}'}(\infty) & = & 
\frac{J_{H}^{2}S}{2} \; f_{{\bf p},{\bf k}',{\bf p}',{\bf k}}
\label{appr_U}
\\
M^{\uparrow}_{{\bf k},{\bf k}',{\bf q},{\bf q}'}(\infty) 
& = & \frac{J_{H}}{2} + \frac{J_{H}^{2}S}{2} \;  
f_{{\bf k},{\bf k}+{\bf q},{\bf k}',{\bf k}'+{\bf q'}} \;,
\label{appr_Mu}
\\
M^{\downarrow}_{{\bf k},{\bf k}',{\bf q},{\bf q}'}(\infty) 
& = & \frac{J_{H}}{2} + \frac{J_{H}^{2}S}{2} \;
f_{{\bf k}'-{\bf q},{\bf k}',{\bf k}-{\bf q}',{\bf k}} \;,
\label{appr_Md}
\end{eqnarray}
where
\begin{equation}
f_{{\bf 1},{\bf 2},{\bf 3},{\bf 4}} = \frac{
\xi^{\uparrow}_{\bf 1} - \xi^{\downarrow}_{\bf 2} 
+ \xi^{\uparrow}_{\bf 3} - \xi^{\downarrow}_{\bf 4}}
{\left( \xi^{\uparrow}_{\bf 1}-\xi^{\downarrow}_{\bf 2}
\right)^{2} + \left( \xi^{\uparrow}_{\bf 3} -
\xi^{\downarrow}_{\bf 4} \right)^{2}} \;.
\label{f_flow}
\end{equation}
Using the standard canonical transformation 
(\ref{standard}) one obtains the same scaling
for the parameters (\ref{appr_xi}-\ref{appr_Md})
but with a different factor $f$ \cite{Golosov-00}
\begin{equation}
f^{(G)}_{{\bf 1},{\bf 2},{\bf 3},{\bf 4}} = 
\frac{1}{2} \; \left( \frac{1}{\xi^{\uparrow}_{\bf 1} 
- \xi^{\downarrow}_{\bf 2}} + \frac{1}{
\xi^{\uparrow}_{\bf 3} - \xi^{\downarrow}_{\bf 4}}
\right) \;.
\label{f_Golosov}
\end{equation}
A general feature of the continuous canonical
transformation is that it derives the effective 
Hamiltonians avoiding any singularities for 
the renormalized energies and interactions 
\cite{Wegner-94,Glazek-94}. For instance, the 
effective retarded interaction between electrons 
in a coupled electron-phonon system has been shown 
to be $-|M^{el-ph}_{\bf q}|^{2} \omega_{\bf q} \left[
\left( \varepsilon_{{\bf k}+{\bf q}}-\varepsilon_{\bf k}
\right)^{2} + \omega_{\bf q}^{2} \right]^{-1} $\cite{el-ph} 
instead of the divergent Fr\"ohlich result 
$|M^{el-ph}_{\bf q}|^{2} \omega_{\bf q} \left[ \left( 
\varepsilon_{{\bf k}+{\bf q}}-\varepsilon_{\bf k} 
\right)^{2} - \omega_{\bf q}^{2} \right]^{-1}$ 
\cite{Frohlich} ($M^{el-ph}_{q}$ denotes the electron
phonon coupling and $\omega_{\bf q}$, $\varepsilon_{\bf k}$
refer to phonon and electron energies respectively). 

A similar situation takes place in the DEX model studied 
here. For small values of $J_{H}S$ (as compared to the 
electron bandwidth) we obtain {\em less divergent} factor 
(\ref{f_flow}) than it has been predicted from the standard 
canonical transformation (\ref{f_Golosov}) found in the Refs 
\cite{Nagaev-98,Golosov-00}. For the limit of large $J_{H}S$ 
both factors (\ref{f_flow},\ref{f_Golosov}) asymptotically 
approach $f \sim - (J_{H}S)^{-1}$ and effectively 
$M^{\sigma} \sim 0$, $U \sim -J_{H}/2$. 

In the limit $l \rightarrow \infty$, the exchange 
interaction (\ref{perturb}) is absent and then one
can roughly estimate the magnon dispersion as
\begin{eqnarray}
\omega_{\bf q} = \frac{1}{N} \sum_{\bf k} 
\left[ M^{\uparrow}_{{\bf k},{\bf k},{\bf q},{\bf q}} 
n_{\bf k}^{\uparrow} - 
M^{\downarrow}_{{\bf k},{\bf k},{\bf q},{\bf q}} 
n_{\bf k}^{\downarrow} \right] \;.
\label{magnon_simple}
\end{eqnarray}
Here the expectation value $n_{\bf k}^{\sigma}=\left<
c_{{\bf k}\sigma}^{\dagger}c_{{\bf k}\sigma} \right>$ 
has a meaning of the Fermi-Dirac distribution function 
of the argument $\xi_{\bf k}^{\sigma}(l=\infty)$. By 
inspecting the expressions (\ref{f_flow},\ref{f_Golosov}) 
one notices that the diagonal terms $f_{{\bf 1},{\bf 2},
{\bf 3}={\bf 1},{\bf 4}={\bf 2}}$ are in both cases 
identical, independently on magnitudes of $J_{H}$ and 
$S$. Effectively, the magnon spectrum becomes  
\begin{eqnarray}
\omega_{\bf q} = \frac{J_{H}}{2N} \sum_{\bf k} 
\left( n_{\bf k}^{\uparrow} - n_{\bf k}^{\downarrow} 
\right) + \frac{J_{H}^{2}S}{2N} \sum_{\bf k} \frac{
n_{\bf k}^{\uparrow}-n_{{\bf k}+{\bf q}}^{\downarrow}}
{\xi^{\uparrow}_{\bf k}-\xi^{\downarrow}_{{\bf k}+{\bf q}}} 
\;. \label{magnon}
\end{eqnarray}
Our lowest order estimation (\ref{magnon}) is thus 
in agreement with predictions based on the standard
canonical transformation \cite{Nagaev-98,Golosov-00}
and other earlier studies of this model 
\cite{Nagaev-69,Furukawa-96,Okabe-97} as well. 
The gapless Goldstone mode $\omega_{{\bf q}=0}=0$ 
(for arbitrary temperature and $J_{H}$) marks 
the spontaneous breaking of rotational symmetry. 

\section{Selfconsistency corrections}  

In this section we take into account the effect of
the terms neglected by us so far in the equation 
(\ref{flow_hamil}). We postulate that the effective
Hamiltonian should be given in the form $H_{0}(l)$ 
(\ref{diag}) with some small correction expressed 
there by $\delta H_{0}(l)$. Following other studies
based on the flow equation method we consider effect
of the higher order interactions by reducing them
to normally ordered form. As can be seen below, 
$\delta H_{0}(l)$ would then be expressed by
the fluctuations around the mean field values, 
which should be small.

i) Correction of the order $O(\hspace{0.2mm}
I\hspace{0.5mm}U\hspace{0.2mm})$ arises from 
the commutator between $\eta$ (\ref{generator})
and the electron electron interaction. It is 
easy to verify that this term can be expressed 
as follows
\begin{eqnarray}
O(\hspace{0.2mm}I\hspace{0.5mm}U\hspace{0.2mm}) 
& = & \frac{1}{N\sqrt{N}} \sum_{{\bf k},{\bf q}} 
\alpha_{{\bf k},{\bf q}}(l) I_{{\bf k},{\bf q}}(l) 
\sum_{{\bf k}',{\bf p}',{\bf p}'',{\bf k}''} 
U_{{\bf k}',{\bf p}',{\bf p}'',{\bf k}''}(l) \; 
\delta_{{\bf k}'+{\bf p}',{\bf k}''+{\bf p}''} 
\nonumber \\ & & \times \;
a_{\bf q}^{\dagger} \left( c_{{\bf p}'\uparrow}^{\dagger}
c_{{\bf k}'\downarrow}^{\dagger} c_{{\bf k}''\downarrow}
c_{{\bf k}\downarrow} \delta_{{\bf k}-{\bf q},{\bf p}''} -
c_{{\bf k}-{\bf q}\uparrow}^{\dagger}c_{{\bf p}'\uparrow}
^{\dagger} c_{{\bf p}''\uparrow}c_{{\bf k}''\downarrow} 
\delta_{{\bf k},{\bf k}'} \right) + \mbox{h.c.} 
\nonumber \\ & = & 
: O(\hspace{0.2mm}I\hspace{0.5mm}U\hspace{0.2mm}) :
\; + \; \frac{1}{N\sqrt{N}} \sum_{{\bf k},{\bf q}}
a_{\bf q}^{\dagger} c_{{\bf k}-{\bf q}\uparrow}
^{\dagger} c_{{\bf k}\downarrow} \sum_{{\bf k}'}
\left[ \alpha_{{\bf k}',{\bf q}}(l) I_{{\bf k}',
{\bf q}}(l) U_{{\bf k}',{\bf k}-{\bf q},{\bf k}'
-{\bf q},{\bf k}}(l) \left( n_{{\bf k}'-
{\bf q}}^{\uparrow} 
\right. \right. \nonumber \\ & & \left. \left.
- n_{{\bf k}'}^{\downarrow} \right) -
\alpha_{{\bf k},{\bf q}}(l) I_{{\bf k},{\bf q}}(l) 
\left( U_{{\bf k},{\bf k}'-{\bf q},{\bf k}'-{\bf q},
{\bf k}}(l) n_{{\bf k}'-{\bf q}}^{\uparrow} - 
U_{{\bf k}',{\bf k}-{\bf q},{\bf k}-{\bf q},
{\bf k}'}(l) n_{{\bf k}'}^{\downarrow} \right)
 \right] + \mbox{h.c.}
\label{OU}
\end{eqnarray}

ii) The other term $O( \hspace{0.2mm} I \hspace{0.2mm} 
M \hspace{0.2mm}$) comes from the commutator between 
$\eta$ and the magnon electron interaction. Its 
contribution to the flow equation (\ref{equation}) 
is
\begin{eqnarray}
O( \hspace{0.2mm} I \hspace{0.2mm} M \hspace{0.2mm})
& = & \frac{1}{N\sqrt{N}} \sum_{{\bf k},{\bf q}} 
\alpha_{{\bf k},{\bf q}}(l) I_{{\bf k},{\bf q}}(l) 
\sum_{{\bf k}',{\bf k}'',{\bf q}',{\bf q}''}
\left[ \delta_{{\bf q},{\bf q}''} \left(M^{\uparrow}
_{{\bf k}',{\bf k}'',{\bf q}',{\bf q}''}(l) 
c_{{\bf k}'\uparrow}^{\dagger} c_{{\bf k}''\uparrow} 
\right. \right. \nonumber \\ & & \; - \left. \left.
M^{\downarrow}_{{\bf k}',{\bf k}'',{\bf q}',{\bf q}''}(l)
c_{{\bf k}'\downarrow}^{\dagger} c_{{\bf k}''\downarrow} 
\right) a_{{\bf q}'}^{\dagger} c_{{\bf k}-{\bf q}
\uparrow}^{\dagger} c_{{\bf k}\downarrow}
 + a_{\bf q}^{\dagger}a_{{\bf q}'}^{\dagger}
a_{{\bf q}''} \left( M^{\uparrow}_{{\bf k}',{\bf k}'',
{\bf q}',{\bf q}''}(l) c_{{\bf k}'\uparrow}^{\dagger} 
c_{{\bf k}\downarrow} \delta_{{\bf k}-{\bf q},{\bf k}''} 
\right. \right. \nonumber \\ & & + \left. \left. 
M^{\downarrow}_{{\bf k}',{\bf k}'',{\bf q}',{\bf q}''}(l) 
c_{{\bf k}-{\bf q}\uparrow}^{\dagger} c_{{\bf k}''\downarrow}
\delta_{{\bf k},{\bf k}'} \right) \right] 
\delta_{{\bf k}'+{\bf q}',{\bf k}''+{\bf q}''} + \mbox{h.c.}
\nonumber \\ & = & : O(IM) : \; + \;
\frac{1}{N\sqrt{N}} \sum_{{\bf k},{\bf q}} 
a_{\bf q}^{\dagger} c_{{\bf k}-{\bf q}\uparrow}^{\dagger} 
c_{{\bf k}\downarrow} \left\{ \alpha_{{\bf k},{\bf q}}(l) 
I_{{\bf k},{\bf q}}(l) \left[ \sum_{{\bf k}'} \left( 
M^{\uparrow}_{{\bf k}',{\bf k}',{\bf q},{\bf q}}(l) 
n_{{\bf k}'}^{\uparrow} 
\right. \right. \right. \nonumber \\ 
& & \left. \left. \left. - 
M^{\downarrow}_{{\bf k}',{\bf k}',{\bf q},{\bf q}}(l) 
n_{{\bf k}'}^{\downarrow} \right) + \sum_{{\bf q}'} 
n_{{\bf q}'} \left( M^{\uparrow}_{{\bf k}-{\bf q},
{\bf k}-{\bf q},{\bf q}',{\bf q}'}(l) + 
M^{\downarrow}_{{\bf k},{\bf k},{\bf q}',
{\bf q}'}(l) \right) \right] 
\right. \nonumber \\ & & + \left. 
\sum_{{\bf q}'} \left[ \left( 1 - n_{{\bf k}-{\bf q}'}
^{\uparrow} + n_{{\bf q}'} \right) M^{\uparrow}_{{\bf k}
-{\bf q},{\bf k}-{\bf q}',{\bf q},{\bf q}'}(l)
\alpha_{{\bf k},{\bf q}'}(l) I_{{\bf k},{\bf q}'}(l)
\right. \right. \nonumber \\ & & + \left. \left. 
\left( n_{{\bf k}-{\bf q}'}^{\downarrow}+n_{{\bf q}'} 
\right) M^{\downarrow}_{{\bf k}-{\bf q}',{\bf k},
{\bf q},{\bf q}-{\bf q}'}(l) \alpha_{{\bf k}-{\bf q}',
{\bf q}-{\bf q}'}(l) I_{{\bf k}-{\bf q}',{\bf q}
-{\bf q}'}(l) \right] \right\} + \mbox{h.c.}
\label{OM}
\end{eqnarray}
We introduced here the expectation value for the 
magnon's number operator $n_{\bf q}=\left< 
a_{\bf q}^{\dagger} a_{\bf q} \right>$, which can
by approximately taken as the Bose-Einstein 
distribution function of the magnon energy 
$\omega_{\bf q}$. 

From now on, we put the both contributions 
(\ref{OU},\ref{OM}) into the $\delta H_{0}(l)$ 
part of the Hamiltonian and neglect there the 
normally ordered parts $:O(IU):$, $:O(IM):$
which are supposed to be small. We neglect 
also a contribution to the flow equation which 
might arise from the commutator $[\eta,\delta H_{0}]$. 
This is the only simplification we need in order to 
close the set of flow equations for the Hamiltonian 
$H_{0}(l)+H_{int}(l)$. By inspecting the structure 
of expressions (\ref{OU},\ref{OM}) we conclude that 
the flow equations (\ref{flow_xi}-\ref{flow_Mdown}) 
remain unchanged. The only quantity affected directly 
is the exchange coupling constant $I_{{\bf k},{\bf q}}(l)$. 
The revised flow equation (\ref{flow_I}) is now given by
\begin{eqnarray}
\frac{dI_{{\bf k},{\bf q}}(l)}{dl} & = &  -  
\alpha_{{\bf k},{\bf q}}^{2}(l) I_{{\bf k},{\bf q}}(l) 
+ \alpha_{{\bf k},{\bf q}}(l) I_{{\bf k},{\bf q}}(l) 
\left\{ \frac{1}{N} \sum_{{\bf k}'} \left[ \left( 
U_{{\bf k},{\bf k}',{\bf k}',{\bf k}}(l) -
M^{\uparrow}_{{\bf k}',{\bf k}',{\bf q},{\bf q}}(l) 
\right) n_{{\bf k}'}^{\uparrow} 
\right. \right. \nonumber \\ & - & \left. \left. \left( 
U_{{\bf k}',{\bf k}-{\bf q},{\bf k}-{\bf q},{\bf k}'}(l) 
- M^{\downarrow}_{{\bf k}',{\bf k}',{\bf q},{\bf q}}(l) 
\right) n_{{\bf k}'}^{\downarrow} \right]
-\frac{1}{N} \sum_{{\bf q}'} n_{{\bf q}'} \left( 
M^{\uparrow}_{{\bf k}-{\bf q},{\bf k}-{\bf q},{\bf q}',
{\bf q}'}(l) 
\right. \right. \nonumber \\ & + & \left. \left.
M^{\downarrow}_{{\bf k},{\bf k},{\bf q}',
{\bf q}'}(l) \right) \right\} +
\frac{1}{N} \sum_{{\bf k}'} \alpha_{{\bf k}',{\bf q}}(l)
I_{{\bf k}',{\bf q}}(l) U_{{\bf k}',{\bf k}-{\bf q},
{\bf k}'-{\bf q},{\bf k}}(l)  \left(
n_{{\bf k}'-{\bf q}}^{\uparrow} - n_{{\bf k}'}^{\downarrow} 
\right) 
\nonumber \\ & - &
\frac{1}{N} \sum_{{\bf q}'} \left[ \left( 1 - 
n_{{\bf k}-{\bf q}'}^{\uparrow} + n_{{\bf q}'} \right) 
M^{\uparrow}_{{\bf k}-{\bf q},{\bf k}-{\bf q}',{\bf q},
{\bf q}'}(l) \alpha_{{\bf k},{\bf q}'}(l) 
I_{{\bf k},{\bf q}'}(l) 
\right. \nonumber \\ & + & \left. \left(
n_{{\bf k}-{\bf q}'}^{\downarrow} + n_{{\bf q}'} 
\right) M^{\downarrow}_{{\bf k}-{\bf q}',{\bf k},{\bf q},
{\bf q}-{\bf q}'}(l) \alpha_{{\bf k}-{\bf q}',{\bf q}-{\bf q}'}(l) 
I_{{\bf k}-{\bf q}',{\bf q}-{\bf q}'}(l) \right] \;. 
\label{flow_Isc}
\end{eqnarray}

We studied numerically the system of coupled flow equations
(\ref{flow_xi}-\ref{flow_Mdown},\ref{flow_Isc}) solving them 
selfconsistently via the Runge Kutta algorithm. Since the model 
parameters such as $U_{{\bf k},{\bf q},{\bf p},{\bf k}+{\bf q}
-{\bf p}}$ and $M^{\sigma}_{{\bf k},{\bf q},{\bf p},{\bf k}-
{\bf q}+{\bf p}}$ depend on three momenta it is a rather 
cumbersome task to study the 3 or even 2 dimensional systems. 
For a demonstration we therefore used the one dimensional 
tight binding lattice with its initial dispersion 
$\varepsilon_{\bf k}(l=0)=-2t \cos{ka}$. From now on, 
we set the lattice constant $a=1$ and choose the initial 
bandwidth as a unit $W=4t \equiv 1$ (flow parameter $l$ 
will be expressed in units of $W^{-2}$).

We discretized the first Brillouin zone with a mesh of
200 equally distant points. Starting from the initial
conditions (\ref{condition1},\ref{condition2}) we computed 
iteratively the renormalized model parameters using 
a following scheme $y(l+\delta l)=y(l)+y'(l)\delta l$ 
where $y'(l)\equiv dy(l)/dl$ and it is given by one of 
the flow equations (\ref{flow_xi}-\ref{flow_Mdown},
\ref{flow_Isc}) for a corresponding parameter $y$. 
Increment of the flow parameter was taken $\delta l=
0.0001$ for $l \leq 0.1$ and $\delta l=0.001$ for 
$l \in (0.1;1.0)$. 

Figure 1 shows how the model parameters evolve with an 
increasing $l$. Practically, already from $l=0.3$ these 
quantities start to saturate at their asymptotic values. 
The exchange interaction disappears very fast, whereas 
the magnon - electron interaction is reduced roughly 
10 times. There occurs some renormalization of 
$\varepsilon^{\downarrow}_{\bf k}$ electron energies 
and a simultaneous induction of a relatively strong 
electron -- electron attractive interactions. 

\begin{figure}
\centerline{\epsfxsize=8cm \epsfbox{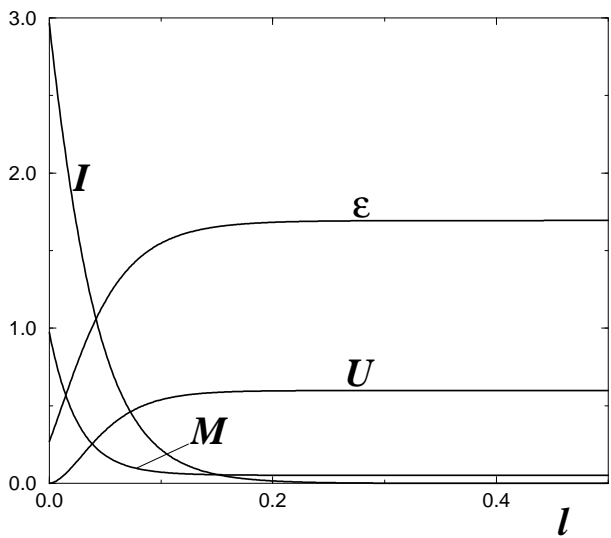}}
\vspace{0.5cm}
\caption{Evolution of the DEX model parameters with respect
to a varying flow parameter $l$ at $T=0$ and hole
concentration $0.3$. Individual lines correspond to 
the following quantities: $ I \equiv \frac{1}{N^{2}} 
\sum_{{\bf k},{\bf q}} |I_{{\bf k},{\bf q}}|^{2}$, 
$\varepsilon \equiv \frac{1}{N} \sum_{\bf k} 
({\varepsilon_{\bf k}}^{\downarrow})^{2}$, 
$U \equiv \frac{1}{N^{3}} \sum_{{\bf k},{\bf p},{\bf q}} 
|U_{{\bf k},{\bf p},{\bf q},{\bf k}+{\bf p}-{\bf q}}|^{2}$ 
and $M \equiv \frac{1}{N^{3}} \sum_{{\bf k},{\bf p},{\bf q}} 
|M_{{\bf k},{\bf p},{\bf q},{\bf k}-{\bf p}+{\bf q}}
^{\downarrow}|^{2}$.}
\end{figure}

\begin{figure}
\centerline{\epsfxsize=13cm \epsfbox{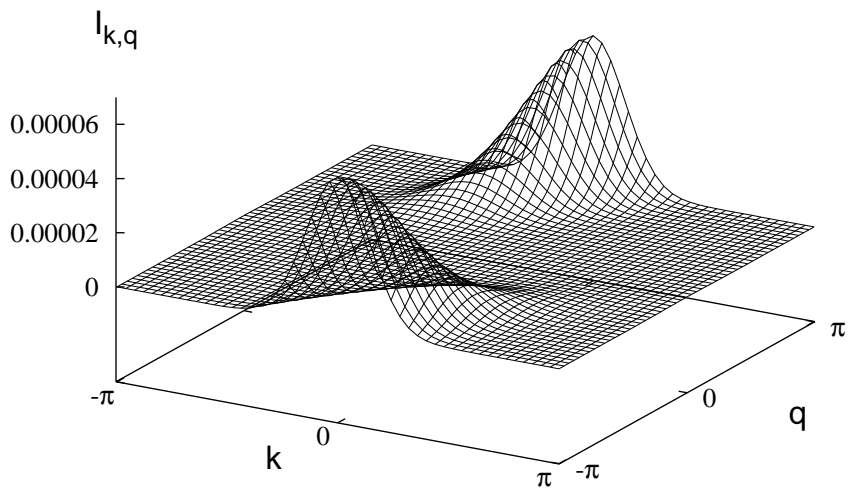}}
\vspace{0.5cm}
\caption{Plot of the normalized exchange coupling 
$I_{\bf k q}(l=1)/I_{\bf kq}(l=0)$. This interaction
is weak enough to be considered as negligible.}
\end{figure}

Figure 2 illustrates what is left of the exchange 
interaction at $l=1$. The highest values of 
$I_{{\bf k},{\bf q}}$ are $7 \times 10^{-5}$
of the initial interaction $J_{H}\sqrt{S/2}$. 
Such small magnitude is in our opinion negligible 
and thereof we represent below the effective model 
parameters (formally corresponding to $l=\infty$) 
through their values obtained at $l=1$.

Let us next have a brief look on the electron part. 
In figure 3 we plot the dispersion for spin $\sigma=
\downarrow$ electrons obtained from the selfconsistent 
solution of the flow equation (\ref{flow_xi}). Effective
bandwidth becomes of the order $\sim 0.83$ of the initial 
bare bandwidth $W$. This renormalization is weaker
than a result of the standard canonical transformation 
\cite{Nagaev-98,Golosov-00}, which gave the scaling 
factor $\sim 0.64$. As concerns the effective mass
we notice a similar tendency. The whole band of 
$\sigma=\downarrow$ electrons drifts away from 
the partly occupied band of $\sigma=\uparrow$ electrons
increasing the gap between them which initially was
$J_{H}S$. The flow equation method gives a somewhat
smaller shift than the standard canonical transformation.

\begin{figure}
\centerline{\epsfxsize=8cm \epsfbox{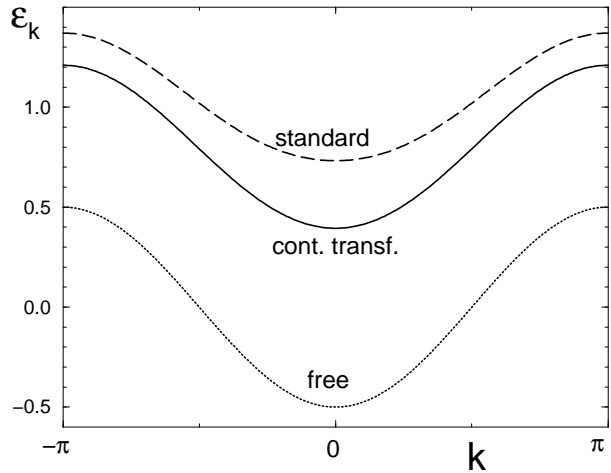}}
\vspace{0.5cm}
\caption{
The effective dispersion for spin $\sigma = \downarrow$ 
electrons obtained from a selfconsistent solution of the 
flow equations (solid line) and via the standard single 
step transformation (dashed line). The dotted curve shows 
the initial bare dispersion $\varepsilon_{\bf k}=-2t
\cos{ka}$.}
\end{figure}

Figure 4 shows a strength of the induced electron -- electron
interaction for the BCS channel $U_{{\bf k},-{\bf k},-{\bf q},
{\bf q}}$ (scattering between the electron pairs of the total 
zero momentum). This interaction is again weaker by almost 
30 $\%$ from the corresponding magnitude determined by 
the standard transformation. It should be stressed here 
that electron -- electron interactions may eventually play 
important role in a low energy physics of the DEX model 
only when hole concentration $1-n$ is close to zero. 
Otherwise, the scattering processes between particles
originating from so much separated ($\sim J_{H}S$) electron
bands would not become efficient.

\begin{figure}
\centerline{\epsfxsize=13cm \epsfbox{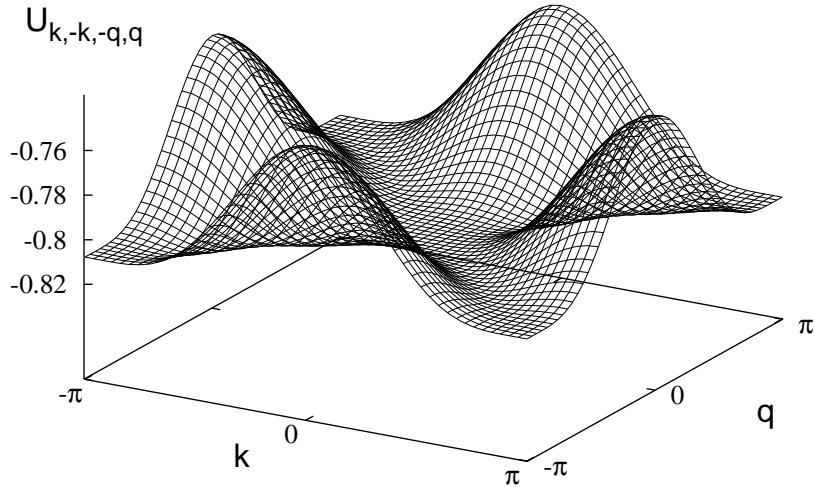}}
\vspace{0.5cm}
\caption{The effective interaction between electrons
in the zero momentum BCS channel, i.e.\ $U_{{\bf k},-{\bf k},
-{\bf q},{\bf q}} c^{\dagger}_{{\bf k}\downarrow} 
c^{\dagger}_{-{\bf k}\uparrow} c_{-{\bf q}\uparrow}
c_{{\bf q}\downarrow}$. Other elements of the potential 
$U_{{\bf k},{\bf p},{\bf q},{\bf k}+{\bf p}-{\bf q}}$
take on average the same values as shown in this picture
(around $-0.8$).}
\end{figure}

\begin{figure}
\centerline{\epsfxsize=8cm \epsfbox{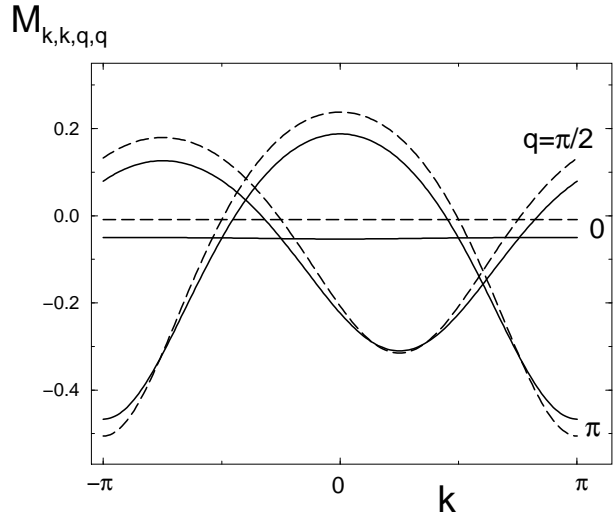}}
\vspace{0.5cm}
\caption{Potential of the electron -- magnon 
interaction $M^{\uparrow}_{{\bf k},{\bf k},{\bf q},
{\bf q}}$ obtained from a selfconsistent solution 
of the flow equations (solid lines) and from 
the standard canonical transformation (dashed 
curves). We used three representative values
for the momentum {\bf q} as marked on a right 
corner of the figure.}
\label{fig5}
\end{figure}

If the initial Hunds coupling $J_{H}$ is large and
hole doping is not close to zero then the low energy
physics of the effective Hamiltonian $H(l=\infty)$ 
is mainly determined from the electron -- magnon 
interaction. This interaction is characterized by
two coupling constants $M^{\sigma}_{{\bf k},{\bf p},
{\bf q},{\bf k}-{\bf p}+{\bf q}}$ which, according
to our numerical estimation, can vary between
$-0.5$ and $0.5$ depending on the momenta {\bf q},
{\bf k}, {\bf p} and only sligthly on temperature $T$
and concentration $n$. In general, in this method 
we find that absolute values of both coupling constants 
$M^{\sigma}$ become reduced by $10$ to $20 \%$ 
as compared to the result of the standard 
transformation (given in equation (\ref{appr_Mu})
with the $f^{(G)}$ factor (\ref{f_Golosov})) .
Figure \ref{fig5} shows this effect on the example 
of the diagonal part $M^{\uparrow}_{{\bf k},{\bf k},
{\bf q},{\bf q}}$ (notice, that in the approximate
solution discussed in section III there was no 
difference between both methods for such diagonal
elements).

A natural expectation is that reduction of the
electron - magnon interaction would in a consequence
affect such quantities like magnon spectrum,
the life-time of these quasiparticles and finally
also the Curie critical temperature. It is not
a scope of this paper to study all these effects
(we mainly intend to present this new technique 
for a derivation of the effective Hamiltonian).
To get some insight we present below a rough
estimation of the magnon dispersion based on
the lowest order formula (\ref{magnon_simple}).
Higher order corrections are straightforward
to carry out but we believe they would not alter
anything in this context.

\begin{figure}
\centerline{\epsfxsize=10cm \epsfbox{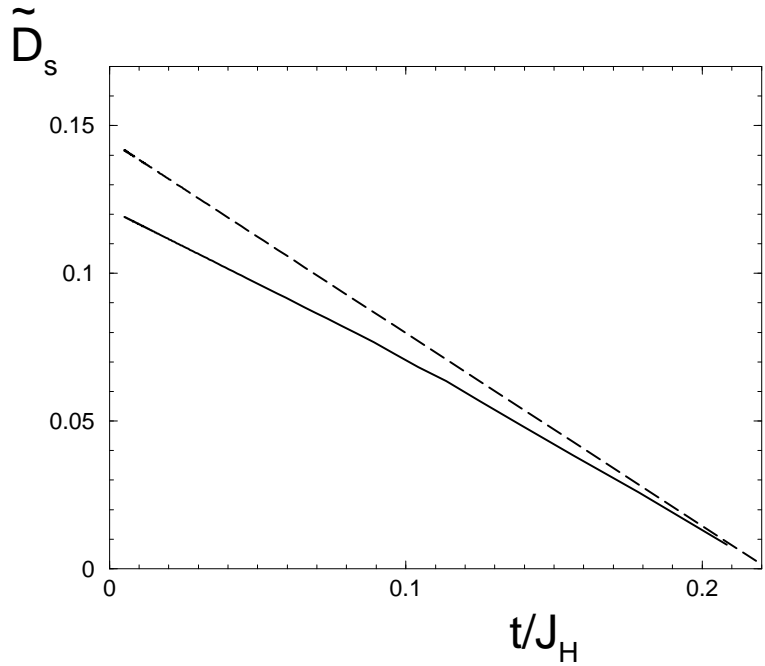}}
\vspace{0.5cm}
\caption{Spin stiffness $\tilde{D}_{s} \equiv 2SD_{s}/t$ 
as a function of the inverse Hund's coupling for
$T=0$ and hole concentration $0.3$. Solid line shows
a result obtained from the flow equation method
and the dashed one refers a standard estimation
based on equation (\ref{magnon}).}
\label{fig6}
\end{figure}

\begin{figure}
\centerline{\epsfxsize=10cm \epsfbox{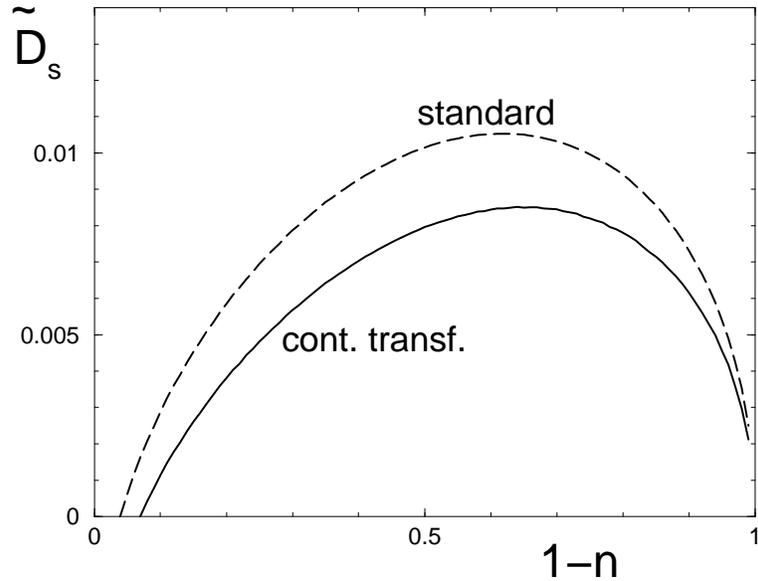}}
\vspace{0.5cm}
\caption{Spin stiffness $\tilde{D}_{s}$ as a function 
of hole concentration for $J_{H}/W=2$, $T=0$ obtained 
from the flow equation method (solid curve) and 
the single step transformation (dashed line).}
\label{fig7}
\end{figure}

In the lowest order perturbative determination of
the magnon dispersion $\omega_{\bf q}$ we simply 
need only the diagonal parts $M^{\sigma}_{{\bf k},
{\bf k},{\bf q},{\bf q}}$ and at low temperatures
main contribution comes from $\sigma=\uparrow$. 
In the long wavelength limit ${\bf q} \rightarrow 0$
for $\omega_{\bf q}$ one gets a parabolic momentum 
dependence with the spin stiffness coefficient $D_{s}
=\lim_{{\bf q} \rightarrow 0} \omega_{\bf q}/q^{2}$.
Using the standard expression (\ref{magnon})
Furukawa \cite{Furukawa-96} gave an explicit
formula for $D_{s}$ at low $T$ and studied $D_{s}$
with respect to $J_{H}$ and the hole concentration
$p=1-n$ (where $n=\sum_{{\bf k},\sigma} 
<c^{\dagger}_{{\bf k}\sigma} c_{{\bf k}
\sigma}>$). For a comparison we show in 
the figures \ref{fig6} and \ref{fig7}
our results. 

Due to a discussed above reduction of the electron 
-- magnon coupling (see figure \ref{fig5}) we notice 
a partial softening of the spin stiffness in a whole 
regime of carrier concentration. Such a softening 
takes place mostly for the strong Hunds coupling. 
Going towards the limit of small $J_{H}$ this effect
becomes less pronounced. In particular, for $J_{H}=
2W$ and hole concentration $p=0.3$ our result for
$D_{s}$ is almost $20 \%$ smaller as compared to
the standard result reported by Furukawa 
\cite{Furukawa-96}.

This tendency for a softening of the spin stiffness 
(especially in a limit of large $J_{H}$) agrees 
qualitatively with the results reported recently 
by Shannon and Chubukov \cite{Shannon-00} who 
introduced a novel large $S$ expansion scheme for
systems with strong Hund's coupling. Authors showed
that quantum effects caused a relative softening
of spin-wave modes at the zone center. There is 
a rich literature where authors report differences
between the Heisenberg cosine dispersion (for 
$\omega_{\bf q}$) and the actual dispersion found 
from the calculations for DEX model 
\cite{Wurth-98,Furukawa-96,Furukawa-99}.
In those papers a relative flattening of the 
dispersion $\omega_{\bf q}$ near the zone
boundaries has been found. Our results follow 
closely the same behavior.

\vspace{1cm}
At the end of this section we would like to make
an effort to analyze in more detail the effect of 
some neglected terms $:O(\hspace{0.2mm}I\hspace{0.5mm}
U\hspace{0.2mm}):$ and $:O(\hspace{0.2mm}I\hspace{0.5mm}
M\hspace{0.2mm}):$. As an illustration, let us consider 
one of possible contributions to $\delta H_{0}(l)$
\begin{eqnarray}
\delta H^{(1)}_{0}(l) & = & \frac{1}{N\sqrt{N}}
\sum_{{\bf q},{\bf k},{\bf p}',{\bf p}''} 
V_{{\bf q},{\bf k},{\bf p}',{\bf p}''}(l) 
: a_{\bf q}^{\dagger} c_{{\bf k}-{\bf q}\uparrow}
^{\dagger} c_{{\bf p}'\uparrow}^{\dagger} 
c_{{\bf p}''\uparrow} c_{{\bf k}+{\bf p}'-{\bf p}''
\downarrow}: 
\label{corr1}
\end{eqnarray}
taken from $:O(IU):$ (\ref{OU}). Its initial ($l=0$)
amplitude is of course $V_{{\bf q},{\bf k},{\bf p}',
{\bf p}''}(l=0)=0$. The flow equation corresponding
to this potential is given by
\begin{eqnarray}
\frac{d V_{{\bf q},{\bf k},{\bf p}',{\bf p}''}(l)}{dl} 
= - \alpha_{{\bf k},{\bf q}}(l) I_{{\bf k},{\bf q}}(l) 
U_{{\bf k},{\bf p}',{\bf p}'',{\bf k}+{\bf p}'
-{\bf p}''}(l) \;.
\label{corr2}
\end{eqnarray}
We checked numerically that the potential $V$ of the 
interaction (\ref{corr1}) is negative. On average, its
value is $1/N^{4}\sum_{{\bf k},{\bf q},{\bf p}',{\bf p}''}
V_{{\bf q},{\bf k},{\bf p}',{\bf p}''} \simeq -0.4$ which 
is close to strength of the electron -- magnon interactions
$<M^{\sigma}>$. To be specific we show in figure 8 this
potential in two channels: 
$V_{{\bf 0},{\bf k},-{\bf k},{\bf q}}(l)
a_{\bf 0}^{\dagger} c_{{\bf k}\uparrow}
^{\dagger} c_{-{\bf k}\uparrow}^{\dagger}
c_{-{\bf q}\uparrow} c_{{\bf q}\downarrow}$
(top picture) and 
$V_{{\bf q},{\bf k},-{\bf k},-{\bf k}}(l)
a_{\bf q}^{\dagger} c_{{\bf k}-{\bf q}\uparrow}
^{\dagger} c_{-{\bf k}\uparrow}^{\dagger}
c_{-{\bf k}\uparrow} c_{{\bf k}\downarrow}$
(bottom picture).
\begin{figure}
\centerline{\epsfxsize=13cm \epsfbox{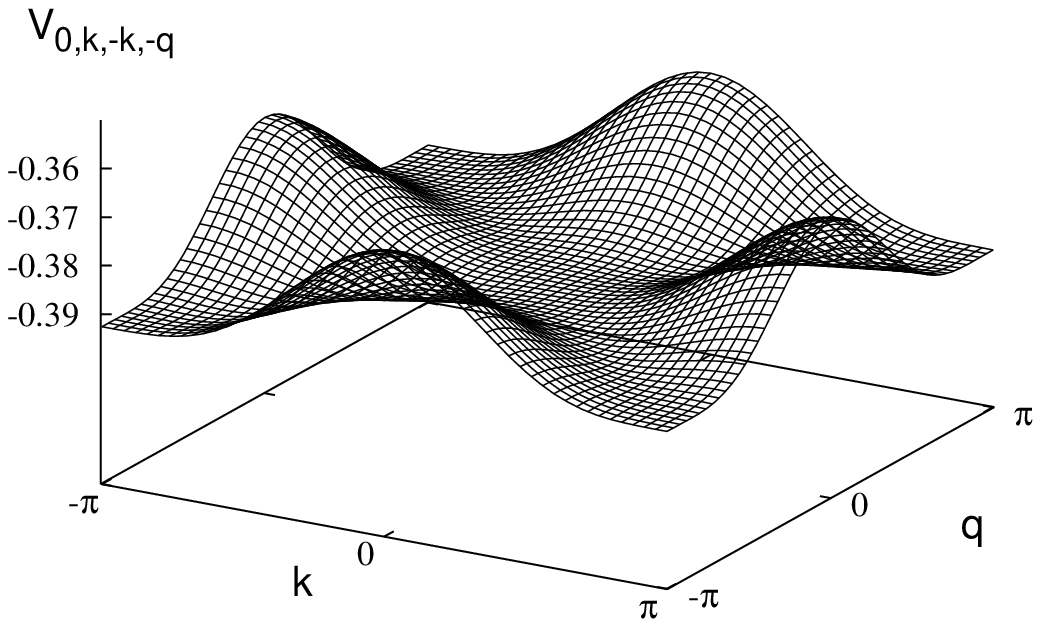}}
\centerline{\epsfxsize=13cm \epsfbox{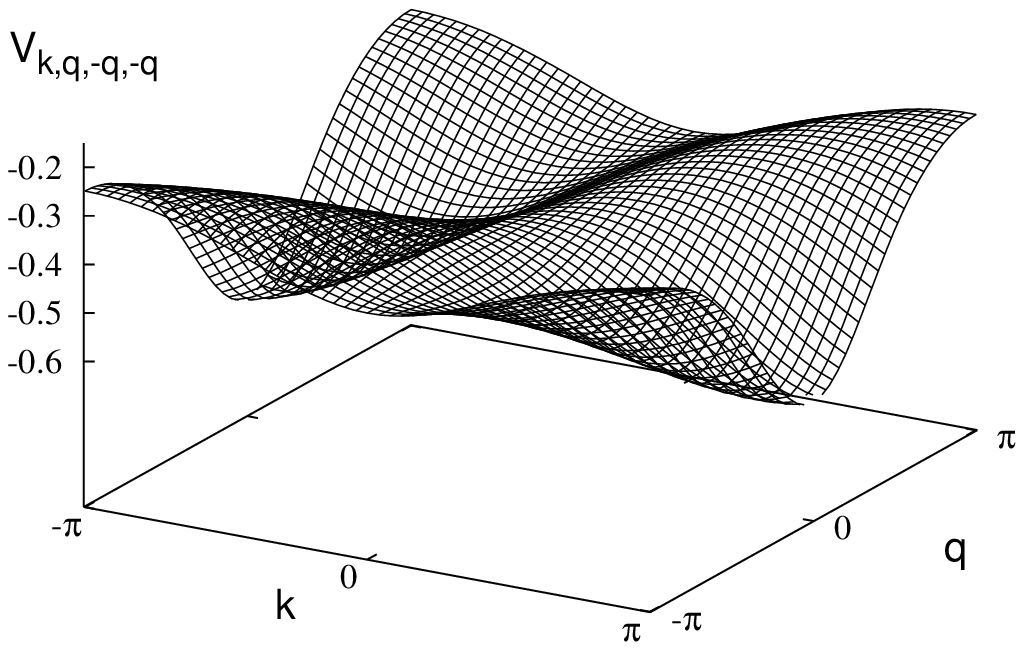}}
\vspace{0.5cm}
\caption{Potential of the many body interaction 
given in the equation (\ref{corr1}) at $T=0$ and 
hole concentration $0.3$}
\end{figure}
Such many body interactions would however be not 
efficient at low temperatures because they engage
electrons from vastly distant bands (similarly 
as the electron -- electron interaction $U$).
Moreover, we have only the normal ordered part
of this interaction entering to the Hamiltonian,
and that should be small. The mean field value
of this type interactions were already included
by us in the flow equation for $I_{{\bf k},{\bf q}}$
(\ref{flow_Isc}).
 
Having interactions shown in (\ref{corr1}), one can 
induce from $[ \eta(l), \delta H_{0}^{(1)}(l) ]$ 
next generation of higher order interactions, like 
for example $c_{{\bf 1}\uparrow}^{\dagger}c_{{\bf 2}
\uparrow}^{\dagger} c_{{\bf 3}\uparrow}c_{{\bf 4}
\uparrow} a_{\bf 5}^{\dagger} a_{\bf 6}$. They can 
be found in the standard canonical transformation as 
well (see $H_{4}'$ in the Ref.\ \cite{Golosov-00}). 
However let us repeat again, that in our case we 
get only the normal ordered forms of all such higher 
order interactions. So, hopefully, their influence
on a physics of the effective Hamiltonian should
be relatively negligible.

\section{Conclusions}

In summary, we formulated continuous canonical 
transformation for the double exchange model, 
eliminating from the initial Hamiltonian the 
exchange interaction term responsible for a 
violation of magnon number. Thus, in a limit
$l \rightarrow \infty $, true magnons are obtained.
Parameters of the effective Hamiltonian are 
determined via the set of the flow equations 
(\ref{flow_xi}-\ref{flow_Mdown}) and (\ref{flow_Isc}).
Structure of the resulting Hamiltonian (\ref{diag})
is simple but the effective model parameters are
computed selfconsistently taking into account
effect of the higher order interactions (what is not 
possible in the standard single step transformation).
These feedback effects are discussed by us on example
of the magnon dispersion. We find a partial softening
of the spin stiffness in a whole regime of hole 
concentration.
 
Further studies are needed to solve the flow equations 
for a realistic 3D version of the DEX model. Another 
important issue not studied here is to extend the 
procedure to capture the damping effects for magnons 
and electrons. As discussed earlier in the context
of different models \cite{spin-boson,Ragawitz-99} one
should study then the {\em flow} not only of the whole
Hamiltonian $H(l)$ but also for the particular operators
like $a^{(\dagger)}_{\bf q}(l)$. Such an investigation
for the DEX model is in progress and the results will be 
reported elsewhere.

\vspace{5mm}
Author kindly acknowledges stimulating discussions 
with G.~Jackeli, J.~Ranninger and K.I.~Wysoki\'nski. 
This work is supported by the Polish Committee of 
Scientific Research under grant No 2P03B 106 18.

\end{document}